\documentclass[3p]{elsarticle}
\usepackage{lipsum}
\makeatletter
\def\ps@pprintTitle{%
 \let\@oddhead\@empty
 \let\@evenhead\@empty
 \def\@oddfoot{}%
 \let\@evenfoot\@oddfoot}
\makeatother

\usepackage{amsmath,amsthm,verbatim,amssymb,amsfonts,graphicx,natbib}

\begin{document}

\begin{frontmatter}

\title{Hadron--Quark Combustion as a Nonlinear, Dynamical System}

\author[cal]{Amir Ouyed\corref{cor2}}
\ead{ahouyedh@ucalgary.ca}

\author[cal]{Rachid Ouyed}
\author[usc]{Prashanth Jaikumar}

\address[cal]{Department of Physics and Astronomy, University of Calgary,
2500 University Drive NW, Calgary, Alberta, T2N 1N4, Canada}
\address[usc]{Department of Physics and Astronomy, California State University Long Beach,
1250 Bellflower Blvd., Long Beach, CA 90840 U.S.A}

\cortext[cor2]{Principal corresponding author}

\begin{abstract}
The hadron--quark combustion front is a system that couples various processes, such as chemical reactions, hydrodynamics, diffusion, and neutrino transport. Previous numerical work has shown that this system is very nonlinear, and can be very sensitive to some of these processes. In~these proceedings, we contextualize the hadron--quark combustion as a nonlinear system, subject to dramatic feedback triggered by leptonic weak decays and neutrino transport. 
\end{abstract}

\end{frontmatter}

\section{Introduction}

The hypothesis of absolutely stable quark matter  \cite{bodmer1971collapsed,witten1984cosmic,terazawa1979tokyo} has very important phenomenological consequences in high energy astrophysics (e.g., \cite{horvath1992nucleation,bombaci2000conversion}).   For example, it quickly became evident that the conversion of a whole neutron star into a quark star could release thermal and mechanical energy that would be in the same order of magnitude than the energy released by a core collapse supernova \cite{benvenuto1989evidence,pagliara2013combustion}. An important reason behind the relevance of  alternative hypothesis such as the one of absolutely stable quark matter is the fact that computational simulations and modelling cannot recreate the energetic events of many explosive astrophysics phenomena.  For example, computational models for core collapse supernovae are still unable to provide robust explosions \cite{muller2016core}. Furthermore, the engines of  even more energetic phenomena, such as super-luminous supernova \cite{abbott2017superluminous} and gamma ray bursts \cite{kumar2015physics} remain elusive. The conversion of hadronic to absolutely stable quark matter could give the extra ``push'' necessary to realize some of these energetic events. Given the potential of this hypothesis of explaining at least in part some of the more mysterious explosive phenomena in astrophysics, the inclusion of this conjecture in models of  explosive astrophysics should remain an active research program (e.g., \cite{welbanks2017simulating}). 

Although simple energetics  reveal the potential of the hypothesis of absolutely stable quark matter,  more sophisticated studies are  necessary to prove whether this conversion would be dynamically significant in, for example, powering a supernova-like explosion \cite{benvenuto1989evidence} or a gamma ray burst \cite{cheng1996conversion}. Given the high densities and the  temperatures of the conversion process, the  most apt framework to study the dynamics of this conversion is hydrodynamics, where  a fluid of neutrons is ``burnt'' into a fluid of quarks \cite{lugones1995strange,herzog2011three}. In the late 1980s, a couple of  papers \cite{olinto1987conversion,benvenuto1989evidence} appeared that pioneered a semi-analytic method of describing the conversion of neutrons into quarks as a hydrodynamic, combustion process. However, the exact equations that govern this process, the reaction--diffusion--advection equations, are quite complicated. These equations couple various processes, such as radiative transfer, chemical reactions, fluid dynamics, and diffusion, forcing these early papers to simplify considerably the equations in order to find a tractable solution. Nevertheless, due to the nonlinear nature of these equations, simplifications that may appear to be minor could actually have dramatic consequences in the equations' solutions.  Nonlinear coupling of various processes could create  dramatic outcomes that could lead to orders of magnitude of difference, in, for example, the speed of the conversion. 

    Later on,  a numerical way of solving directly these reaction--diffusion--advection equations was pioneered by Niebergal et al. \cite{niebergal2010numerical} (hereby Paper I).   Their numerical results lead to large differences to much of the previous semi-analytic work. For example, the calculated burning speed of about 0.002c--0.04c, where c is the speed of light,  was orders of magnitude faster than what was calculated previously by some semi-analytic models. Their results also hinted at important feedback effects that might arise from coupling neutrino transport into the combustion front. In fact, they found, by solving the  semi-analytic, hydrodynamic jump conditions, that, for a given neutrino cooling rate, thermal pressure gradients could slow down the combustion front by orders of magnitude. \mbox{Ouyed et al. \cite{ouyed2017numerical}} (hereby Paper II) later confirmed this initial intuition numerically, by incorporating neutrino transport and electron pressure into the reaction--diffusion--advection equations. The authors of Paper II discovered that feedback effects triggered by various leptonic processes could affect the burning timescale by orders of magnitude.  These two papers showed then that lepton micro-physics are at the very least as important as other parts of the combustion process that have already been deemed important for decades, such as the high density equation of state. The importance of leptons follows from the nonlinear nature of the combustion process---parts of the system that may appear at first glance insignificant may give rise to extreme feedback effects. In the case of neutrinos, we have shown that their omission in simulations would lead to very inadequate results, given that they can dramatically affect the conversion speed. 

    These proceedings will  therefore focus on the importance of lepton micro-physics as a source of nonlinear, feedback effects, by summarizing and contextualizing previously published work, and detailing possible future avenues of research. By micro-physics, we mean the processes that are important at a length-scale of a centimeter, rather than macroscopic processes that appear at the length scale of a compact star (about ten kilometers). Given that, at least to the extent of our knowledge, previous work has never contextualized the issues of hadron--quark combustion using the framework of nonlinear dynamics, we feel that these proceedings could act as a brief introduction to a new way of thinking about the hadron--quark combustion. In particular, we find the concept of feedback loops to be very relevant and illuminating the  micro-physics of the flame. In nonlinear dynamics, a feedback loop implies that the output of a system is fed back into input, creating a circuit of cause and effect that can lead to dramatic consequences. The processes coupled in the combustion front can lead to feedback loops,  where  processes that slow down the burning front could in turn trigger other processes (e.g.,~the magnifying of pressure gradients) that would slow down the burning front further.
        
        We structure these proceedings in the following way. In Section 
        \ref{feedback}, we describe the intricate structure of the flame and the processes that are coupled in it, and how these processes may lead to feedback. In Section \ref{leptons}, we focus on the effect of neutrinos and electrons, which are the source of the main feedback effects described in Paper I and II. In Section \ref{conclusion}, we finish with some concluding remarks.

    \section{Feedback Effects and the Reaction Zone}\label{feedback}

    The reactions that  drive the burning front are:
    
    \begin{equation}\label{gamma1}
u + e^- \leftrightarrow s + \nu_{e},
\end{equation}
\begin{equation}\label{gamma2}
u + e^- \leftrightarrow d + \nu_{e},
\end{equation}
\begin{equation}\label{gamma3}
u + d \leftrightarrow u + s.
\end{equation}

    These reactions are coupled to the reaction--diffusion--advection equations that govern the burning \cite{niebergal2010numerical,ouyed2017numerical}:
\begin{equation}
    \dfrac{\partial n_i}{\partial t} = -\nabla \cdot (n_i v - D_i \nabla n_i) + R_i,
\end{equation}

\begin{equation}\label{momentum}
    \dfrac{\partial (h v)}{\partial t} = -\nabla  (h v \cdot v) - \nabla P,
\end{equation}

\begin{equation}
    \dfrac{\partial s}{\partial t} = -\nabla \cdot (sv) - \frac{1}{T} \sum_i \mu_i \dfrac{d n_i}{d t}+ \frac{1}{T} \dfrac{d \epsilon_{\nu_e}}{d t},
\end{equation}
where the index i runs through the different particle species (u, d, s, $\nu_{e}$). The definition of the  variables are: $n_i$ is number density, $v$ is the fluid velocity,  $\epsilon_{\nu_e}$ is electron neutrino energy density,  $h$ is enthalpy density, $s$ is entropy density, $T$ is temperature, $R_i$ is the reaction source term, $D_i$ is the diffusion coefficient, and $P$ is pressure. 
    
We enforce charge neutrality by equating the electron number density with $n_e=n_u -n_B$. This is a good assumption given that electrons are degenerate and relativistic, so they move with speeds close to the speed of light in order wash out any charge imbalances. 
    
\begin{figure}
\centering 
\includegraphics[width= 0.7\textwidth]{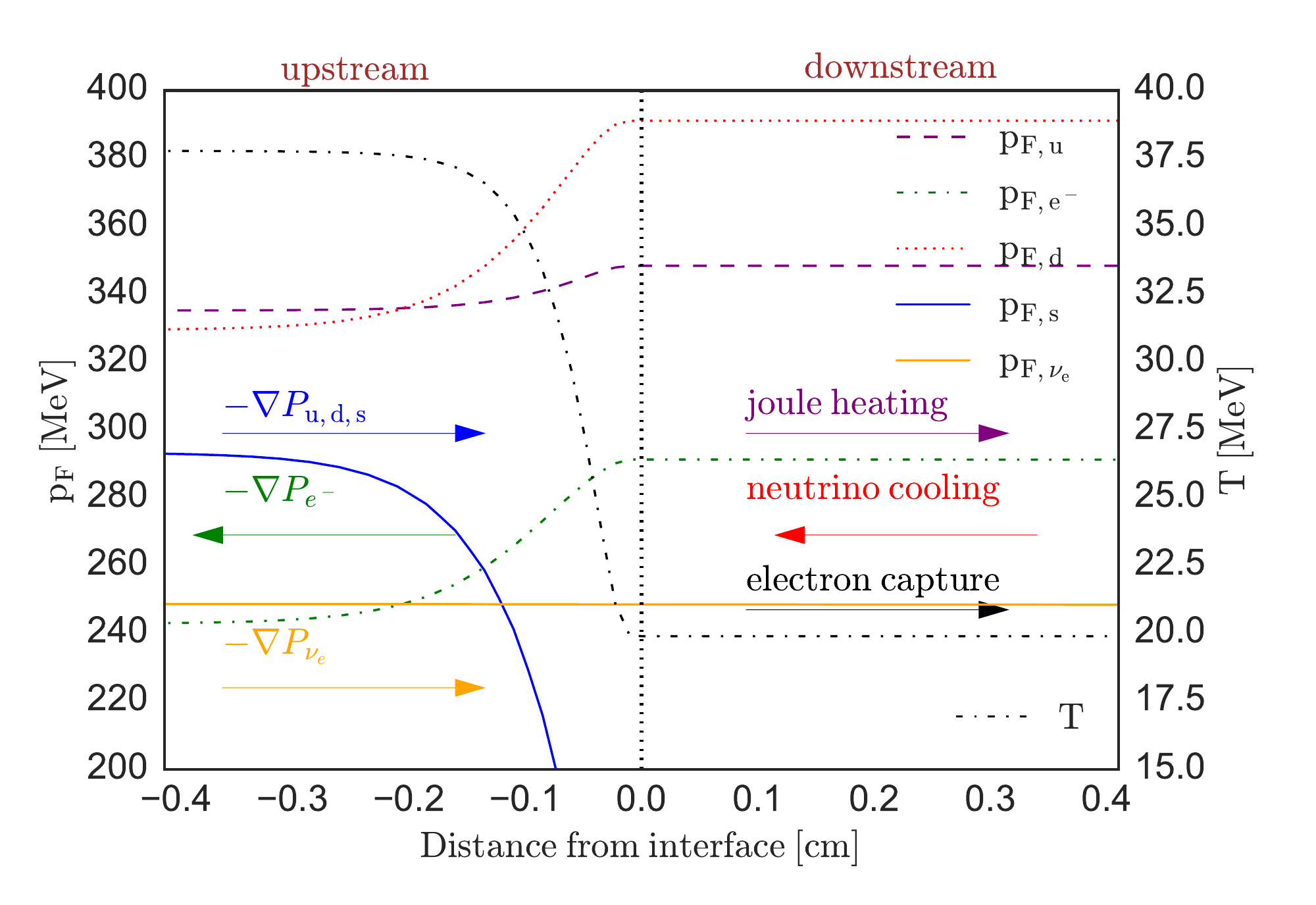}
\includegraphics[width= 0.65\textwidth]{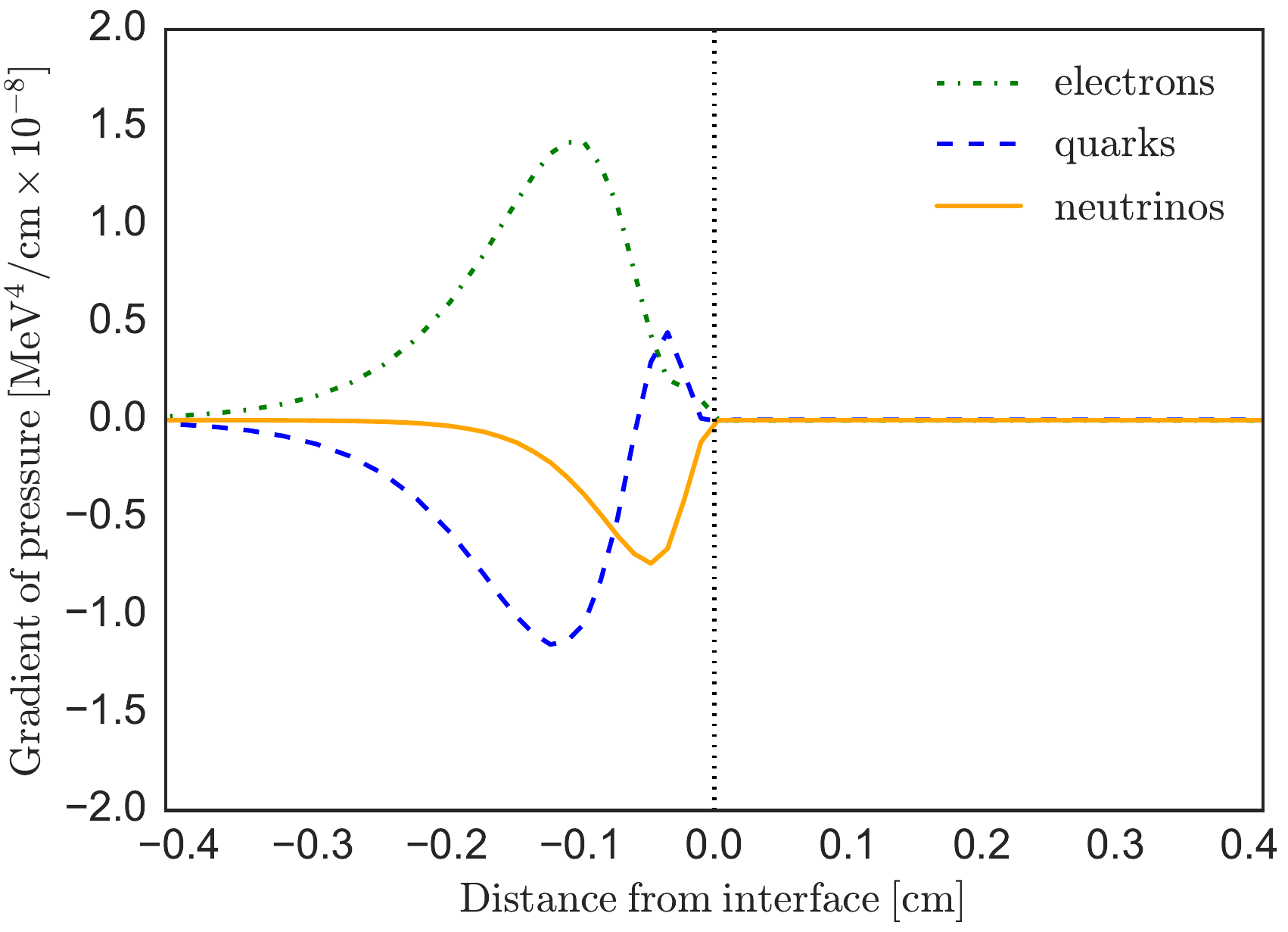}

  \caption{\textbf{Panel a)}: Simulation snapshot of the burning interface. $p_{F,i}$ are the Fermi momenta for particles \emph{i}, and $T$ is the temperature. In both panels, the interface lies at position zero depicted by the vertical line.  The arrows represent the directions of the force vectors and their labels depict the processes that caused them. Upstream is the side behind (left side of the vertical line) the interface, and downstream is the side in front of it (right side of the vertical line). 
   \textbf{Panel b)}: The pressure gradients for the leptons and quarks shown in panel a). Figure and caption were taken from Paper II \cite{ouyed2017numerical}.}
    
  \label{zoner}  
\end{figure}

 These equations lead to a reaction zone that acts as an interface between the two flavoured quark ``fuel'' and the  three flavoured quark ``ash''.  The reaction zone is very complex, given that various particles and processes participate in it. Figure \ref{zoner}, which is a snapshot of numerical simulations performed in Paper II,  is included to illustrate the  complexity of the reaction zone. Panel a) in Figure \ref{zoner} shows the Fermi momenta of various particles and the temperature gradient along the flame, with the various force gradients caused by the processes. Panel b) shows the various pressure gradients caused by the different particles.  Ultimately, the reactions will be constrained by the transport of s-quarks into the fuel, given that the s-quark acts as an ``oxidant'' that triggers the conversion. These transport processes shape the width of the reaction zone, which  is a function of  the nonlinear, hydrodynamic effects related to the distance that fluid velocities  carry the  u- and d-quarks before they decay into s-quarks. Therefore, much of the processes manifest as force vectors that either slow down the transport of s-quarks into the fuel or accelerate it. These processes  may accelerate the burning front or slow it down. We divide the processes along enhancing and quenching, although this division is a simplification because the various processes may be coupled to each other.  Enhancing processes  accelerate the burning front while quenching processes slow it down. We define the burning speed as the derivative of the interface position versus time.

     The enhancing processes are (with more detailed explanations in Paper II): 
 
 \begin{enumerate}

	\item \textbf{Flavor equilibration}:  The conversion of two flavoured quark matter to three flavoured quark matter through the reactions \eqref{gamma1}--\eqref{gamma3} releases binding energy in the form of heat, increasing the temperature behind the front. The increase of temperature stiffens the quark EoS, increasing the pressure behind the front and therefore accelerating the burning speed.

	\item \textbf{Electron capture}: The transformation of electrons into neutrinos through reactions \eqref{gamma1} and \eqref{gamma2} releases binding energy in the form of heat. Higher temperature enhances the pressure behind the interface, which increases the burning speed. 
		
	\item  \textbf{Neutrino pressure}:   Neutrinos deposit momentum into the reaction zone, accelerating the interface into faster speeds.

	\item  \textbf{Loss of lepton number}: Neutrinos, as they diffuse from higher to lower chemical potentials, deposit the chemical potential difference in the form of heat. This heat increases the temperature and therefore enhances the pressure behind the interface. This phenomenon is very similar to what is referred as Joule heating in papers concerning proto--neutron star evolution (e.g., \cite{burrows1986birth}).

	\end{enumerate}

The quenching factors are: 
	\begin{enumerate}

	\item \textbf{Electron pressure}: Electron capture ``eats up'' the electrons behind the interface, generating a large electron gradient (see the electron Fermi momentum distribution in panel a) of Figure \ref{zoner}).  These electron gradients generate a degeneracy pressure that pushes the interface backwards, decelerating the burning front. See panel b) of Figure \ref{zoner} for a graphical representation of the electron pressure gradient.

	\item  \textbf{Neutrino cooling}: Neutrinos that escape from the burning front carry energy away from the reaction zone, which reduces the temperature and therefore the pressure behind the interface.  This quenching effect was first detailed in Paper I. 

    \end{enumerate}
    \vspace{-6pt}

        All these enhancing and quenching processes generate feedback effects.  In the case of the hadron--quark combustion front, positive feedback could be how some  processes that slow down the burning front in turn lead to other processes (e.g., amplification of certain pressure gradients) that would lead to even more deceleration, generating a nonlinear, exponential effect. Although there are probably many types of feedback loops in the reaction zone given its rich couplings  of particles and processes, Papers I and  II focused on the positive feedback generated by leptonic weak interactions, which we will explore in the next section. 

    \section{Leptons and Positive Feedback}\label{leptons}

       Electrons and neutrinos are  crucial components in the combustion system because they can generate dramatic, positive feedback effects. The authors of Paper I were the first to discover a connection between leptons and positive  feedback.  They solved the hydrodynamic jump-conditions for the conversion of two flavoured quark matter to three flavoured quark matter, and parameterized neutrino cooling as a small  temperature reduction in the three flavoured quark ash. They found that, for a very small amount of cooling---for example, a reduction of 0.1 MeV in the temperature---the thermal pressure would reduce dramatically to the point of almost halting the burning interface.

\begin{figure}
\centering 
\includegraphics[width= 0.65\textwidth]{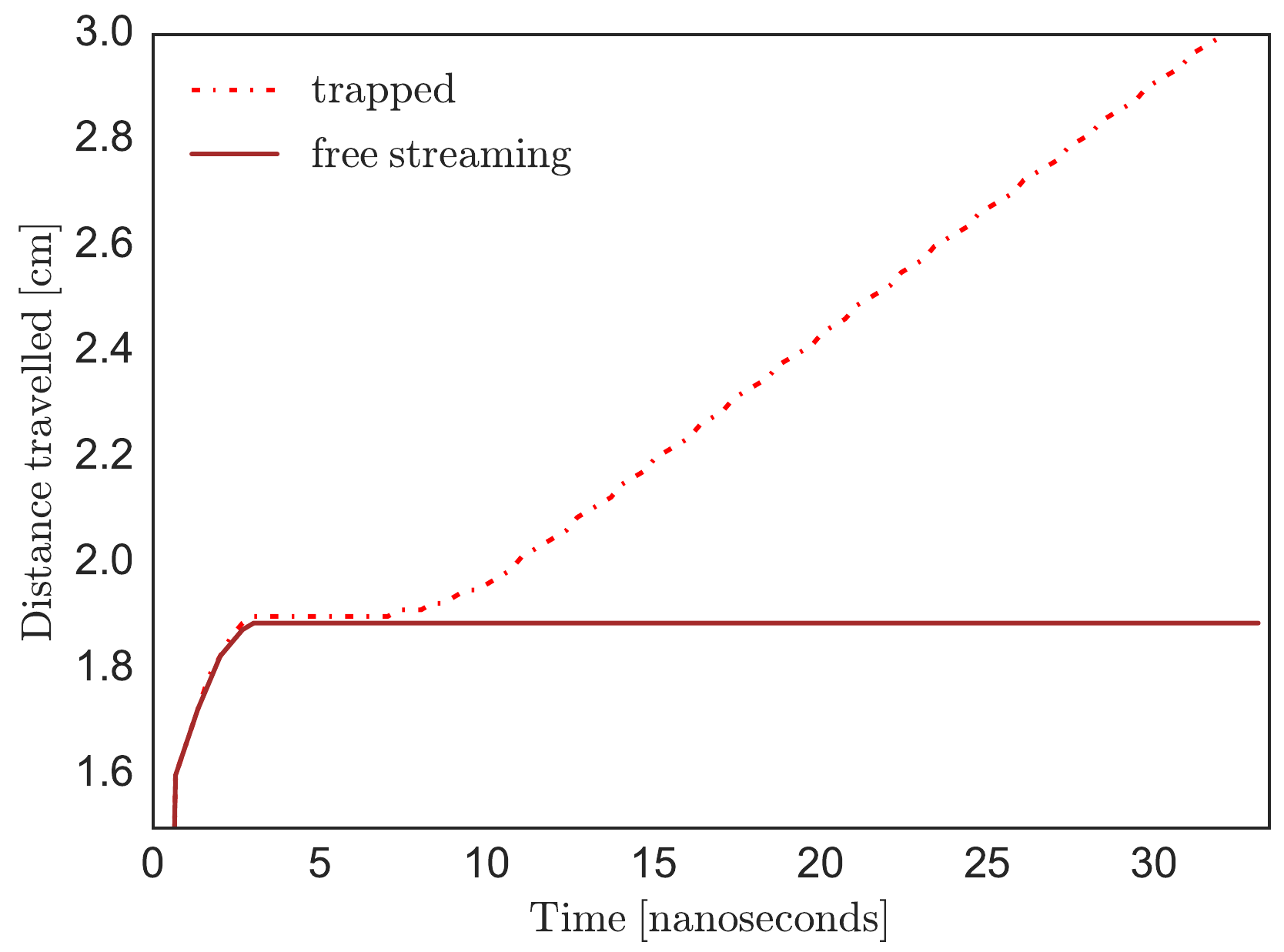}
  \caption{Distance travelled by the combustion front as a function of time from a numerical simulation. The line labeled as ``free streaming'' represents the burning front with neutrinos free streaming, while the line labeled as ```trapped'' plots the burning front with trapped neutrinos.  Notice how the front halts for the remainder of the simulation for the free streaming  case. The thermodynamic parameters for the simulation were an initial temperature of \emph{T} = 20 MeV, an initial lepton fraction of \emph{Y}$_{L}=0.2$, and an initial baryonic density of $n_{{B}}=0.35$ fm$^{-3}$.}
    
  \label{core2}  
\end{figure}

      Paper II discovered more positive feedback effects associated with leptons, and was the first attempt in the literature to incorporate neutrino transport across the reaction zone numerically.  Paper II showed, through a combination of semi-analytic studies and numerical simulations, that the leptons themselves can generate huge pressure gradients that can affect dramatically the burning speed (panel b) in Figure \ref{zoner} ).  A key finding is that the quenching process of electron capture could generate  positive feedback that could slow down the burning front dramatically if the neutrinos are free streaming, to the point that the burning front halts within the timescales of the simulation (Figure \ref{core2}).

      Much of the source of the dramatic lepton feedback lies  in Equation \eqref{momentum} given that the nonlinear momentum is coupled to a lepton degeneracy pressure component in the $\nabla P$ term.  This lepton pressure term in turn is coupled to reaction source terms, entropy evolution and the transport equation of neutrinos.  Given that  velocity varies by various orders of magnitude through the simulation timescale, this equation cannot be linearized (e.g., the $\nabla  (h v \cdot v)$ term in Equation \eqref{momentum} is strongly nonlinear), as the time-dependent fluid velocity is not merely a perturbation oscillating around an equilibrium point, but instead changes dramatically through time by a forcing due to the $\nabla P$ term. In~other words, the burning interface is genuinely a nonlinear system out of equilibrium,  and linearizing the system would eliminate the dramatic feedback loops, generating an inaccurate solution.  
      
      One of the positive pieces of feedback caused by leptons can be described in the following way: processes that slow down the burning front will magnify the electron pressure gradients (see panel b) of Figure \ref{zoner}) that oppose the front, given that it gives reactions  \eqref{gamma1} and \eqref{gamma2} more time to ``eat up'' the electrons behind the front, generating a sharper electron dip. As the electron dip (see panel a) of  Figure \ref{zoner} for a graphical representation of the electron dip) becomes more dramatic and sharper, this in turn will lead to a slower burning front that would induce an even sharper dip, generating a positive feedback effect. In contrast, a faster burning front might move too quickly for flavour equilibration to  catch up  with the interface, weakening the electron pressure gradient that opposes the front. 
      
      Similarly, another positive piece of feedback caused by quenching effects  and associated with leptons is that  slower burning leads to slower reaction rates and therefore less neutrinos being created, which reduces neutrino enhancing processes such as Joule heating and neutrino pressure. As the rate of production of neutrinos decreases, so does the burning speed. 
      
      A key finding is that quenching processes of electron capture could generate positive feedback that could slow down the burning front dramatically if neutrinos are free streaming, to the point that the burning front halts within the timescales of the simulation (Figure \ref{core2}). This is because neutrino pressure and Joule heating counteracts the quenching effects of the electron pressure gradient, and therefore once the neutrinos free stream and therefore do not deposit heat or momentum in the reaction zone, the electron pressure gradient stands unopposed, slowing down the burning front by various orders of magnitude. Only when neutrinos are trapped, such as is the case of Figure 3 in Paper II, will the burning front revive, given that neutrinos deposit momentum and heat that pushes the interface; otherwise, if the neutrinos free stream, the front will remain halted. In Paper II, we showed, both through self-consistent simulations and confirmed it with semi-analytic results, that, if neutrinos free stream, the front will effectively halt due to electron pressure gradients.

    Therefore, a key message of Paper I and II is the realization that the reaction zone of hadron--quark combustion is a nonlinear system that is quite sensitive to its different parts. As an example of this sensitivity, Paper I and II uncovered the fact that leptonic weak interactions generate nonlinearities that could slow down the burning front's speed by various orders of magnitude. 
    
{\em Instabilities in multidimensions:} Paper I and II discovered that  the behavior of the burning front is very sensitive to the distribution of neutrinos and electrons.  It would be interesting to entertain how these lepton sensitivities manifest in more than one dimension. Given that a realistic, multidimensional compact star model would have spatial anisotropies in its lepton distribution, the burning front would probably halt unevenly across its surface, which would lead to  wrinkling.  This would be a novel and alternative channel of turbulence, which would exist alongside other more classical instabilities, such as Rayleigh--Taylor or Kelvin--Helmholtz. Therefore, a multidimensional simulation would be ultimately the decisive factor for unearthing the final fate of a neutron star converting into a quark star. It could be that the instabilities slow down the burning dramatically, or lead to cataclysmic outcomes such as quark core-collapse or supersonic burning speeds that would lead to an explosion.  Analogous cases that demonstrate the significance of multidimensional studies are core-collapse supernova simulations, where  multi-dimensional, hydrodynamic instabilities have turned out to be important for the understanding of the explosion mechanism \cite{muller2016core}. 

The important thing to point out is that the nonlinearities unearthed in one dimension simulations hint that the burning speeds extracted from micro-physical, one-dimensional laminar studies are not the end of it all. Therefore,  the astrophysical implications can only be fully understood with a multidimensional  code.

     \section{Conclusions}\label{conclusion}
     
    We reviewed the recent literature on the micro-physical simulations of  hadron--quark combustion in order to sketch an outline of the reaction zone as a nonlinear system that experiences feedback. Thus, the main objective of these proceedings was to point out that the combustion front, because of its nonlinear nature, can experience significant feedback and coupling between various parts of the system (e.g., quarks, neutrinos, entropy generations), therefore linearizing the problem, and, ignoring certain parts of the system, can generate an inaccurate picture of its behavior.  A key finding is that leptonic weak decays are a key part of this nonlinear system, where the coupling of electrons and neutrinos to entropy generation and hydrodynamics can lead to positive feedback loops that can quench the burning speed almost completely.
    
    Given the nonlinear, dynamical nature of the hadron--quark combustion front, we describe the following common pitfalls in the literature that we attempt to remedy by numerical simulation:
    
    \begin{enumerate}
    \item  Assuming the system is steady-state, in other words, equating all temporal derivatives to zero.
    \item Assuming that the front is in pressure equilibrium, that is, fixing $\nabla P =0$. 
     \item The above two points lead to the cancellation of  the important nonlinearities. Pressure equilibrium and a steady-state momentum make the fluid velocity a constant in space and time. 
    \item Another related pitfall is collapsing the rich structure of the reaction zone into a discontinuity by solving the jump conditions instead of the continuous hydrodynamic equations. This also leads to a steady-state solution, which eliminates the dynamism of the system.

    \end{enumerate}
    
    We must reiterate that the micro-physical simulations reviewed are only in one dimension, and therefore they only offer hints to how these nonlinearities might manifest multidimensionally.  Nonetheless, the nonlinear effects hint that fluid-dynamical instabilities could be induced by the coupling of leptons to the fluid, which was termed as a  deleptonization instability.  This potential deleptonization instabilites, which can only be truly probed with multidimensional simulations, leaves the real timescale of the burning of the whole compact star into a quark star an open question. The~deleptonization instability could slow down the burning of the whole compact star to a matter of hours, or accelerate it towards supersonic detonation that would last less than a millisecond.  Therefore, the next pressing step is to hopefully extend these microphysical simulations into the multidimentional regime, which may unearth new and very interesting nonlinearities. 
    
   Another important question that arises is what more sophisticated and accurate numerical modeling has to offer for the one-dimensional case. For example, in this study, the exact neutrino  Boltzmann transport equations are simplified into  isotropic, energy averaged flux-limited diffusion equations. However, since the diffusion approximation assumes that the neutrinos are thermalized and therefore strongly interacting with matter, a more accurate approximation will make matter more transparent to neutrinos, and therefore exacerbate the positive feedback described in this paper because less neutrino momentum and heat will be deposited in the reaction zone. Another approach that could increase the accuracy of the simulation is higher resolution.  The most recent simulations we ran used a grid of 600 cm with 48,000 zones (size of zone is  dx = 0.0125 cm). However, a higher resolution would probably exacerbate the positive feedback given that the pressure gradients may become sharper due to the reduction of numerical viscosity. Higher order spatial, finite difference schemes (in~this work, we used a third order scheme for advection and second order scheme for diffusion) would reduce numerical viscosity as well. Our work therefore acts as a lower bound to the effect of leptons on the interface, with more sophisticated numerical approaches probably magnifying their effect.

\section*{Acknowledgements}
R.O. is funded by the Natural Sciences and Engineering Research Council of Canada under Grant No. RT731073. P.J. is supported by the U.S. National Science Foundation under Grant No. PHY 1608959.


\begin{thebibliography}{999}
\providecommand{\natexlab}[1]{#1}

\bibitem[Bodmer(1971)]{bodmer1971collapsed}
Bodmer, A.
\newblock Collapsed nuclei.
\newblock {\em Phys. Rev. D} {\bf 1971}, {\em 4},~1601--1606.

\bibitem[Witten(1984)]{witten1984cosmic}
Witten, E.
\newblock Cosmic separation of phases.
\newblock {\em Phys. Rev. D} {\bf 1984}, {\em 30},~272--285.

\bibitem[Terazawa(1979)]{terazawa1979tokyo}
Terazawa, H.
\newblock INS-report, 336 (INS, Univ. of Tokyo); 1989.
\newblock {\em J. Phys. Soc. Japan}, 58(3555):1989, 1979. 

\bibitem[Horvath \em{et~al.}(1992)Horvath, Benvenuto, and
  Vucetich]{horvath1992nucleation}
Horvath, J.; Benvenuto, O.; Vucetich, H.
\newblock Nucleation of strange matter in dense stellar cores.
\newblock {\em Phys. Rev. D} {\bf 1992}, {\em 45},~3865--3868.

\bibitem[Bombaci and Datta(2000)]{bombaci2000conversion}
Bombaci, I.; Datta, B.
\newblock Conversion of neutron stars to strange stars as the central engine of
  gamma-ray bursts.
\newblock {\em Astrophys. J. Lett.} {\bf 2000}, {\em 530},~L69.

\bibitem[Benvenuto and Horvath(1989)]{benvenuto1989evidence}
Benvenuto, O.; Horvath, J.
\newblock Evidence for strange matter in supernovae?
\newblock {\em Phys. Rev. Lett.} {\bf 1989}, {\em 63},~716--719.

\bibitem[Pagliara \em{et~al.}(2013)Pagliara, Herzog, and
  R{\"o}pke]{pagliara2013combustion}
Pagliara, G.; Herzog, M.; R{\"o}pke, F.K.
\newblock Combustion of a neutron star into a strange quark star: The neutrino
  signal.
\newblock {\em Phys. Rev. D} {\bf 2013}, {\em 87},~103007.

\bibitem[M{\"u}ller(2016)]{muller2016core}
M{\"u}ller, B.
\newblock The Core-Collapse Supernova Explosion Mechanism.
\newblock {\em Proc. Int. Astron. Union} {\bf 2016},
  {\em 12},~17--24.

\bibitem[Abbott \em{et~al.}(2017)Abbott, Cooke, Curtin, Joudaki, Katsianis,
  Koekemoer, Mould, Tescari, Uddin, and Wang]{abbott2017superluminous}
Abbott, T.; Cooke, J.; Curtin, C.; Joudaki, S.; Katsianis, A.; Koekemoer, A.;
  Mould, J.; Tescari, E.; Uddin, S.; Wang, L.
\newblock Superluminous Supernovae at High Redshift.
\newblock {\em Publ. Astron. Soc. Aust.} {\bf
  2017}, {\em 34}.

\bibitem[Kumar and Zhang(2015)]{kumar2015physics}
Kumar, P.; Zhang, B.
\newblock The physics of gamma-ray bursts \& relativistic jets.
\newblock {\em Phys. Rep.} {\bf 2015}, {\em 561},~1--109.

\bibitem[Welbanks \em{et~al.}(2017)Welbanks, Ouyed, Koning, and
  Ouyed]{welbanks2017simulating}
Welbanks, L.; Ouyed, A.; Koning, N.; Ouyed, R.
\newblock Simulating Hadronic-to-Quark-Matter with Burn-UD: Recent work and
  astrophysical applications.
\newblock  \emph{J. Phys. Conf. Ser.}  \textbf{2017}, 
\emph{ 861}, ~012008.

\bibitem[Cheng and Dai(1996)]{cheng1996conversion}
Cheng, K.; Dai, Z.
\newblock {Conversion of neutron stars to strange stars as a possible origin of
  $\gamma$-ray bursts.}
\newblock \mbox{{\em Phys. Rev. Lett.}} {\bf 1996}, {\em 77},~1210--1213.

\bibitem[Lugones and Benvenuto(1995)]{lugones1995strange}
Lugones, G.; Benvenuto, O.
\newblock Strange matter equation of state and the combustion of nuclear matter
  into strange matter in the quark mass-density-dependent model at T > 0.
\newblock {\em Phys. Rev. D} {\bf 1995}, {\em 52},~1276--1280.

\bibitem[Herzog and R{\"o}pke(2011)]{herzog2011three}
Herzog, M.; R{\"o}pke, F.K.
\newblock Three-dimensional hydrodynamic simulations of the combustion of a
  neutron star into a quark star.
\newblock {\em Phys. Rev. D} {\bf 2011}, {\em 84},~083002.

\bibitem[Olinto(1987)]{olinto1987conversion}
Olinto, A.V.
\newblock On the conversion of neutron stars into strange stars.
\newblock {\em Phys. Lett. B} {\bf 1987}, {\em 192},~71--75.

\bibitem[Niebergal \em{et~al.}(2010)Niebergal, Ouyed, and
  Jaikumar]{niebergal2010numerical}
Niebergal, B.; Ouyed, R.; Jaikumar, P.
\newblock Numerical simulation of the hydrodynamical combustion to strange
  quark matter.
\newblock {\em Phys. Rev. C} {\bf 2010}, {\em 82},~062801.

\bibitem[Ouyed \em{et~al.}(2017)Ouyed, Ouyed, and Jaikumar]{ouyed2017numerical}
Ouyed, A.; Ouyed, R.; Jaikumar, P.
\newblock Numerical simulation of the hydrodynamical combustion to strange
  quark matter in the trapped neutrino regime.
\newblock {\em Phys. Lett. B} {\bf 2017}, \emph{777}, 184--190.

\bibitem[Burrows and Lattimer(1986)]{burrows1986birth}
Burrows, A.; Lattimer, J.M.
\newblock The birth of neutron stars.
\newblock {\em Astrophys. J.} {\bf 1986}, {\em 307},~178--196.

\end{thebibliography}
\end{document}